\documentclass{aa}
\usepackage[utf8]{inputenc}
\usepackage{graphicx}        % For eps figures, newer & more powerfull
\usepackage[colorlinks=True,citecolor=blue,linkcolor=blue]{hyperref}

\usepackage{natbib}         % For citations: redefine \cite commands
\usepackage{amssymb}        % useful mathemati cal symbols
\usepackage{amsmath}

\usepackage{color} % For color text: \color command
\usepackage{MR_AA}
\usepackage[normalem]{ulem}
\graphicspath{{fig/}}

\title{An improved asteroseismic age of the rapid rotator\\ Altair from TESS
data\thanks{This paper includes data collected by the TESS mission. Funding for
the TESS mission is provided by the NASA's Science Mission Directorate. Data of
the light curve shown in Fig. 2 are available in electronic form at the CDS via
anonymous ftp to cdsarc.u-strasbg.fr (130.79.128.5) or via http://cdsweb.u-strasbg.fr/cgi-bin/qcat?J/A+A/.}}
\author{Michel Rieutord\inst{1} \and Daniel R. Reese\inst{2} \and Joey~S.~G. Mombarg\inst{1} \and Stéphane Charpinet\inst{1}}
\date{\today}
\institute{
IRAP, Universit\'e de Toulouse, CNRS, UPS, CNES,
14, avenue \'{E}douard Belin, F-31400 Toulouse, France\\
\email{Michel.Rieutord@irap.omp.eu}
\and
LESIA, Observatoire de Paris, Université PSL, CNRS, Sorbonne Université, Univ. Paris Diderot,                             Sorbonne Paris Cité, 5 place Jules Janssen, 92195 Meudon, France
}

\titlerunning{An improved asteroseismic age of Altair from TESS data}
\authorrunning{M. Rieutord et al.}

\begin{document}

\abstract
%context
{Understanding the effects of rotation in stellar evolution is key to modelling early-type stars, half of which have equatorial velocities over 100~km/s. The nearby star Altair is an example of such fast-rotating stars, and furthermore, it has the privilege of being modelled by a detailed 2D concordance model that reproduces most of its observables.}
%aims
{The aim of this paper is to include new asteroseismic frequencies to improve our knowledge of Altair, especially its age.}
%Methods
{We processed images of Altair obtained during July 2022 by the Transiting Exoplanet Survey Satellite using the halo photometry technique to obtain its light curve over this observation period.}
%Results
{By analysing the light curve, we derived a set of 22 new frequencies in the oscillation spectrum of Altair and confirmed 12 previously known frequencies. Compared with model predictions, we could associate ten frequencies with ten axisymmetric modes. This identification is based on the modelled visibility of the modes. Moreover, nine of the modelled frequencies can be adjusted to simultaneously match their corresponding observed frequencies, once the core hydrogen mass fraction of the concordance model is set to $X_{\rm core}/X_{\rm ini}\simeq0.972$, with $X_{\rm ini}=0.739$. Using the combined results of a 1D \MESA model computing the pre-main sequence and a 2D time-dependent \ester model computing the main sequence, we find that this core hydrogen abundance sets the age of Altair to 88$\pm$10~Myrs, which is slightly younger than previous estimates.}
%Conclusions
{}


\maketitle

\section{Introduction}

Within 10 parsecs of the Solar System, Altair is the star with the
fastest rotation rate. Its equatorial velocity projected on the
line of sight (V sin$i$) reaches 240~km/s, while interferometric
observations show that the inclination of the rotation axis on the
line of sight is $i=50.65\pm1.23^\circ$ \citep{bouchaud+20}, leading
to an equatorial velocity of 310~km/s. As a consequence of this high
equatorial velocity, Altair is strongly distorted by the centrifugal
force, thus making it a privileged target for imaging by interferometry
\citep{vanbelle+01,domiciano+05,petersonetal06a,monnier+07,bouchaud+20,spalding+22}.
Actually, this star also shows $\delta$-Scuti type oscillations,
{as revealed by the work of \cite{buzasietal05} using data collected
in 1999 by the Wide Field Infrared Explorer (WIRE) satellite star
tracker \citep{buzasietal05}. This detection  was later
confirmed by \cite{ledizes+21} with data from the Microvariability and
Oscillations of STars (MOST) satellite \citep{walker+03},
which observed Altair in 2007, 2011, 2012, and 2013. These first space
observations revealed 15 eigenfrequencies of the star. Soon after, using
spectroscopic time series, \cite{rieutord+23} showed the presence of
prograde surface perturbations associated with internal gravito-inertial
waves.} This wealth of data from interferometry and the first seismic
frequencies of \cite{buzasietal05} motivated \cite{bouchaud+20} to
attempt a 2D modelling of this star using the public domain \ester
code \citep{ELR13,RELP16}. Hence, \cite{bouchaud+20} managed to design
the first concordance model of Altair, and based on this model, it
turned out that this star is $\sim$100~Myrs old and has a mass of
1.86$\pm$0.03~\msun.

To progress further in the understanding of the properties of this star, an improved modelling is needed, and new data are of course most welcome. In this article, we present the results of our processing of data recently collected by the Transiting Exoplanet Survey Satellite \citep[TESS;][]{ricker+15} and investigate its first implications on Altair's modelling. Altair was in the field of view of TESS from 10 July to 5 August 2022. However, it is one of the brightest stars in the sky, and therefore, it saturated TESS' detectors. To circumvent this problem, we resorted to halo photometry as described by \cite{white+17} in order to exhibit Altair's light curve.

In the following, we present the way we derived Altair's light curve from TESS data (Section 2) and give the list of frequencies resulting from its analysis (Section 3). We then focus on the identification of the most visible modes using models of adiabatic oscillations based on the \ester and \topcode codes (Section 4). Finally, we discuss the implications of the results of the estimated age of Altair (Section 5), and conclusions follow.

\section{Preparation of the data}

With an apparent visual magnitude of 0.76 (TESS magnitude is 0.58), the photon flux from Altair largely saturates the pixel where the star falls. To circumvent this difficulty, one may concentrate on pixels that received the overflowing electrons at a level below saturation or those fed by the scattered light, namely those on the halo that has been generated by the generous light flux from the star. \cite{white+17} (W17 hereafter) have shown that this halo still contains information on the flux variations of the star. Provided that these variations are on a timescale much longer than the jitter of the instrument or its random noise, W17 showed that one can recover the star light curve by selecting pixels that minimise the flux variations during the monitoring of the star. As shown in this work, the flux at time $n$ may be written as

\begin{equation}
    f_n=\sum_{i=1}^M w_ip_{in},
\end{equation}
where $w_i$ is the weight of the $i$th pixel and $p_{in}$ is the flux in this pixel at time $n$. The weights satisfy

\begin{equation}
    w_i>0 \andet \sum_{i=1}^M w_i =1\; ,
\end{equation}
where $M$ is the number of selected pixels. As discussed by W17, the term $w_i$ can be determined by minimising a `normalized first-order total variation (TV) of fluxes' defined by 
\begin{equation}
{\rm    TV} = \frac{\sum_{n=1}^N |f_n-f_{n-1}|}{\sum_{n=1}^N f_n}
\label{TV},
\end{equation}
where $N$ is the total number of frames. We refer the reader to the article of W17 for a detailed presentation and discussion of the halo photometry method.

\begin{figure}
\includegraphics[width=0.9\linewidth]{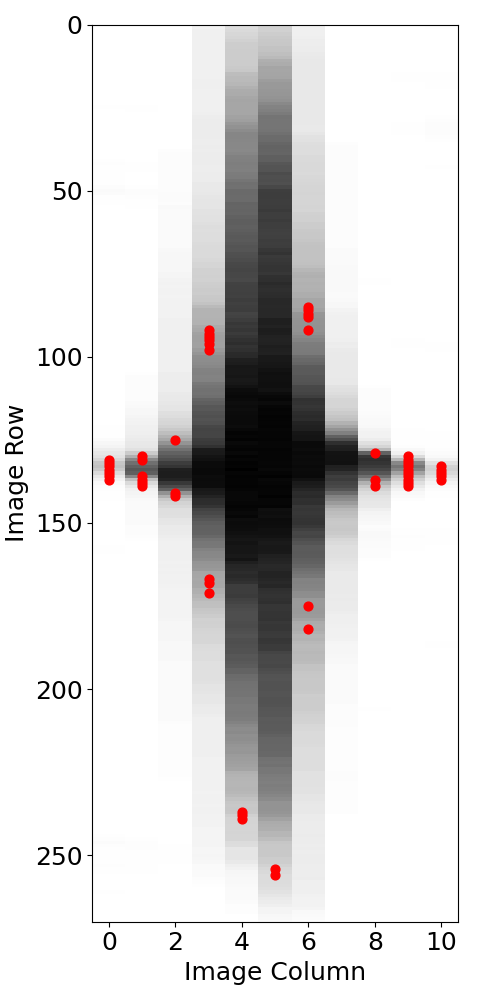}
\caption{Thumbnail image of Altair as obtained with TESS. The red dots mark the position of the selected pixels used to monitor the flux of the star.}
\label{act_pixel}
\end{figure}

From the Mikulski Archive for Space Telescopes (MAST), we downloaded the target pixel file associated with Altair\footnote{tess2022190063128-s0054-0000000070257116-0227-s\_tp.fits} on sector 54. It contains a sequence lasting 26.23 days starting from MJD 59770.40 with three interruptions of 4.18, 1.19, and 2.90 days. {The larger gaps correspond to the time when the Earth was close to the field of view of the camera recording Altair (Camera 1), while the small gap of 1.19 days is due to data download at perigee \citep{TESS22}.} The sampling cadence was 120 seconds, so we processed 18890 images, out of which we extracted N=13720 images of 430$\times$11 pixels with usable data.

As shown in Fig.~\ref{act_pixel}, the images are extremely elongated in order to capture the overflowing electrons from the saturated pixels. {As a first step in the processing, we removed the fluctuating background using the {\tt source extractor} software \citep{barbary15,bertin_arnouts96}}. We then evaluated the saturation of the images at a level of $\sim148,000$~ADU, from which we decided to mask all pixels with a flux level higher than 120,000~ADU. Similarly, we also masked pixels with a level less than 1000~ADU. All the rows beyond index 270 were also dismissed. These thresholds were chosen by varying them so as to optimise the final signal-to-noise ratio of the detected modes. These thresholds led to a selection of 230 valid pixels.

The minimisation process of the TV function (Eq.~\ref{TV}) was done using the `Sequential Least Squares Programming' of the scipy library \citep{scipy2020} following W17. The minimisation process led to the selection of 55 pixels that actually monitor the flux of Altair. These pixels are shown in Fig.~\ref{act_pixel} (red dots) along with a thumbnail image. We cannot exclude the possibility that the selected pixels are contaminated by some field stars since Altair lies at a low galactic latitude (-8$\fdg91$) and the stellar background density is high. However, inspection of the TESS field around Altair showed that neighbouring stars are all fainter than mag. 9.5, that is at least 3,300 times fainter, allowing us to believe that any contamination should be less than a percent. {This level of contamination is acceptable for our purpose of extracting oscillation frequencies.}

\begin{figure}
\includegraphics[width=\linewidth]{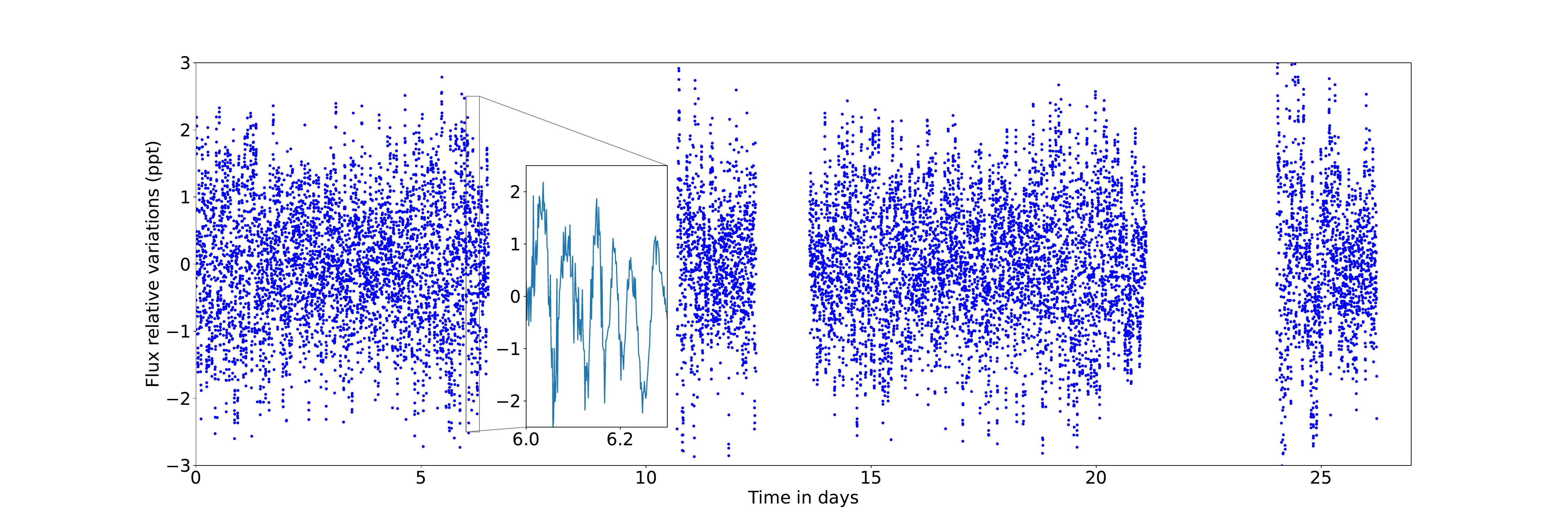}
\caption{Light curve resulting from the processing of TESS images. The general sampling period was 2 minutes. The time origin corresponds to MJD 59770.40.}
\label{LC}
\end{figure}

By using the selected pixels, removing a few outliers, and subtracting a
polynomial fit of degree 9 designed for each of the four sub-sequences,
we obtained the neat light curve shown in Fig.\ref{LC}. The polynomial
fit subtraction increases the signal-to-noise ratio, but it can also
suppress frequencies (typically) less than one cycle per day. Filtered and
unfiltered data used to compute the light curve are available in electronic form
(see title footnote).

\begin{figure}
\includegraphics[width=\linewidth]{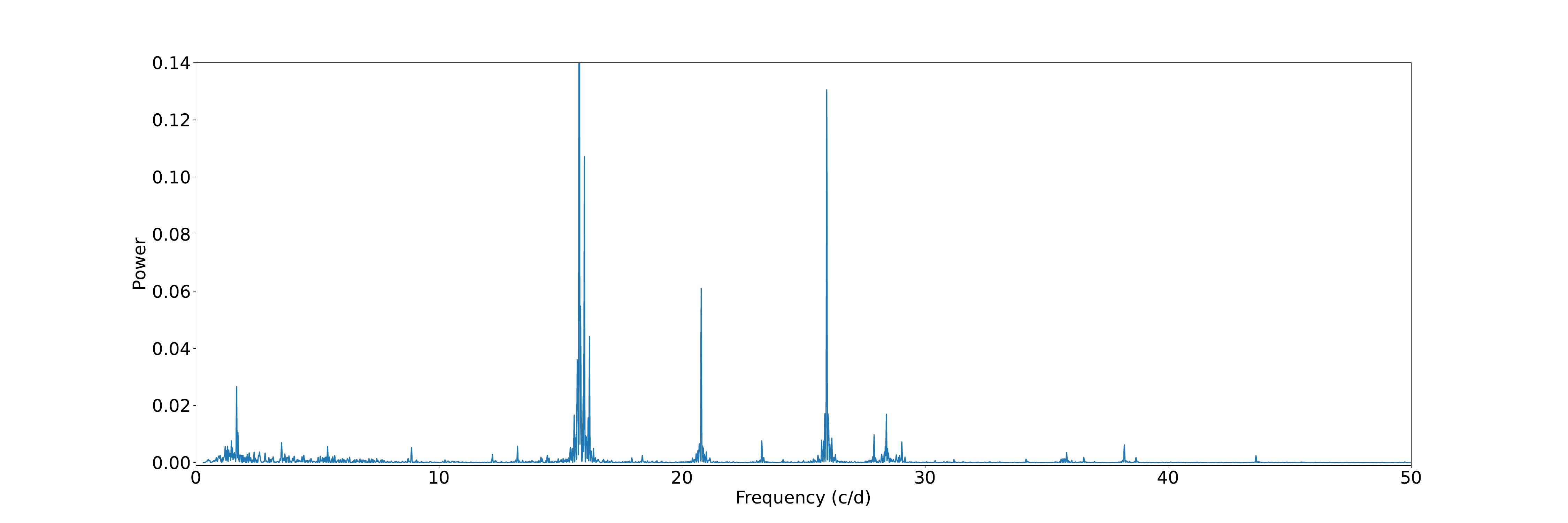}
\caption{Altair spectrum as derived from the light curve of Fig.~\ref{LC}. The power unit is arbitrary.}
\label{SP}
\end{figure}

We then analysed the light curve by computing the classical Lomb-Scargle periodogram (LSP), pre-whitening the signal and adjusting a linear combination of sinusoids with a non-linear least-square optimisation. We used the \FELIX code \citep{charpinet2010,zong2016} to process the light curve. The oscillation spectrum of Altair is shown in Fig.~\ref{SP}, and the extracted frequencies are given in Table~\ref{tab_freq}. We adopted a conservative detection threshold of S/N=5, following the discussion of \citet{zong2016}, with noise being evaluated locally around each frequency as the median value of the residual in the LSP after subtraction of all significant peaks. 

\footnotesize
\begin{table}
\caption{Frequencies, $f_{n}$, and their fitted parameters. A minimum S/N=5 was imposed.}
\centering
\begin{tabular}{lcccccc}
\hline\hline
Id. &   & Frequency  & $\sigma_f$      &  A& $\sigma_A$ & S/N \\
    &   & (c/d)      & (c/d)           &  (ppm)       & (ppm)    &     \\
\hline
$f_{6}$  & & $1.6693$   & $0.0018$     &  $229$   & $20$    &  $11.6$   \\
$f_{11}$  & & $3.5203$   & $0.0031$    &  $121$   & $18$    &  $6.9$   \\
$f_{9}$  & & $5.4148$   & $0.0027$     &  $129$   & $16$    &  $7.9$   \\
$f_{17}$  & & $5.4832$   & $0.0037$    &  $93$   & $16$     &  $5.7$   \\
$f_{15}$  & & $8.8664$   & $0.0030$    &  $98$   & $14$     &  $6.9$   \\
$f_{19}$  & & $12.1978$   & $0.0036$   &  $74$   & $13$     &  $5.8$   \\
$f_{16}$  & & $13.2289$   & $0.0026$   &  $95$   & $12$     &  $8.0$   \\
$f_{23}$  & & $14.1887$   & $0.0036$   &  $65$   & $11$     &  $5.8$   \\
$f_{24}$  & & $14.4547$   & $0.0037$   &  $64$   & $11$     &  $5.7$   \\
$f_{1}$  & & $15.76789$   & $0.00030$  &  $731$   & $10$    &  $70.8$   \\
$f_{4}$  & & $15.98242$   & $0.00073$  &  $293$   & $10$    &  $28.7$   \\
$f_{5}$  & & $16.19097$   & $0.00091$  &  $234$   & $10$    &  $23.1$   \\
$f_{27}$  & & $17.9326$   & $0.0036$   &  $53.6$   & $9.2$  &  $5.8$   \\
$f_{20}$  & & $18.3647$   & $0.0027$   &  $69.7$   & $9.0$  &  $7.7$   \\
$f_{3}$  & & $20.78755$   & $0.00051$  &  $342.9$   & $8.3$ &  $41.1$   \\
$f_{10}$  & & $23.2791$   & $0.0013$   &  $128.5$   & $7.8$ &  $16.5$    \\
$f_{34}$  & & $24.1551$   & $0.0038$   &  $42.7$   & $7.7$  &  $5.6$     \\
$f_{31}$  & & $25.4012$   & $0.0033$   &  $47.9$   & $7.5$  &  $6.4$     \\
$f_{2}$  & & $25.95061$   & $0.00031$  &  $507.5$   & $7.6$ &  $67.2$   \\
$f_{29}$  & & $27.5606$   & $0.0030$   &  $51.0$   & $7.3$  &  $7.0$    \\
$f_{8}$  & & $27.9016$   & $0.0011$    &  $140.5$   & $7.3$ &  $19.4$   \\
$f_{13}$  & & $28.3610$   & $0.0013$   &  $117.8$   & $7.2$ &  $16.3$  \\
$f_{7}$  & & $28.40638$   & $0.00079$  &  $191.3$   & $7.2$ &  $26.5$  \\
$f_{21}$  & & $28.8124$   & $0.0023$   &  $67.0$   & $7.2$   & $9.3$   \\
$f_{12}$  & & $29.0424$   & $0.0013$   &  $119.6$   & $7.2$  & $16.6$   \\
$f_{32}$  & & $29.2514$   & $0.0032$   &  $47.5$   & $7.2$   & $6.6$   \\
$f_{33}$  & & $31.1858$   & $0.0033$   &  $44.1$   & $7.0$   & $6.3$   \\
$f_{30}$  & & $34.1522$   & $0.0028$   &  $50.3$   & $6.7$   & $7.5$   \\
$f_{25}$  & & $35.6018$   & $0.0023$   &  $61.7$   & $6.6$   & $9.3$   \\
$f_{18}$  & & $35.8214$   & $0.0015$   &  $91.6$   & $6.6$   & $13.9$   \\
$f_{26}$  & & $36.5285$   & $0.0025$   &  $55.5$   & $6.5$   & $8.6$   \\
$f_{14}$  & & $38.1975$   & $0.0013$   &  $104.1$   & $6.4$  & $16.2$   \\
$f_{28}$  & & $38.6778$   & $0.0025$   &  $53.0$   & $6.3$   & $8.4$    \\
$f_{22}$  & & $43.6137$   & $0.0020$   &  $65.5$   & $6.1$   & $10.7$   \\
\hline
\end{tabular}
\label{tab_freq}
\end{table}
\normalsize

\begin{figure}
\includegraphics[width=0.95\linewidth]{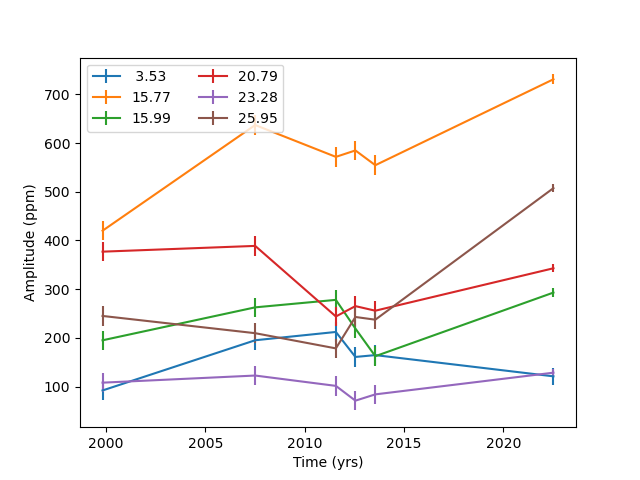}
\caption{Amplitude variations over the years of the most prominent modes.}
\label{Ampl}
\end{figure}

\section{Results}

In view of the frequencies and amplitudes extracted from the light curve, we note that Altair's excited eigenmodes are coherent over a long time base. Mode frequencies found in the previous data, {going back to the} years 1999, 2007, 2011, 2012, and 2013 \cite[see][]{ledizes+21}, are found again in 2022, except for one frequency at 2.58 c/d and two frequencies less than 1 c/d. The two latter frequencies, if they are indeed intrinsic, have most likely been removed by the subtraction of the polynomial fit, as mentioned above. The disappearance of the peak at 2.58~c/d observed in 1999, 2012, and 2013, suggests that this is a beating frequency resulting from the superposition of modes with frequencies $f_2,f_3,f_7,f_{10}$, which show an average difference of 2.54~c/d. The more accurate data collected by TESS removed this frequency. However, the new data confirm four of the six new frequencies exhibited by \cite{ledizes+21}, namely $f_6,f_{27},f_{12}$, and $f_{18}$, and add many other new frequencies. The oscillation spectrum of Altair now contains at least 34 frequencies that need be explained.

\begin{figure*}[t!]
\centering
\includegraphics[width=1.\linewidth]{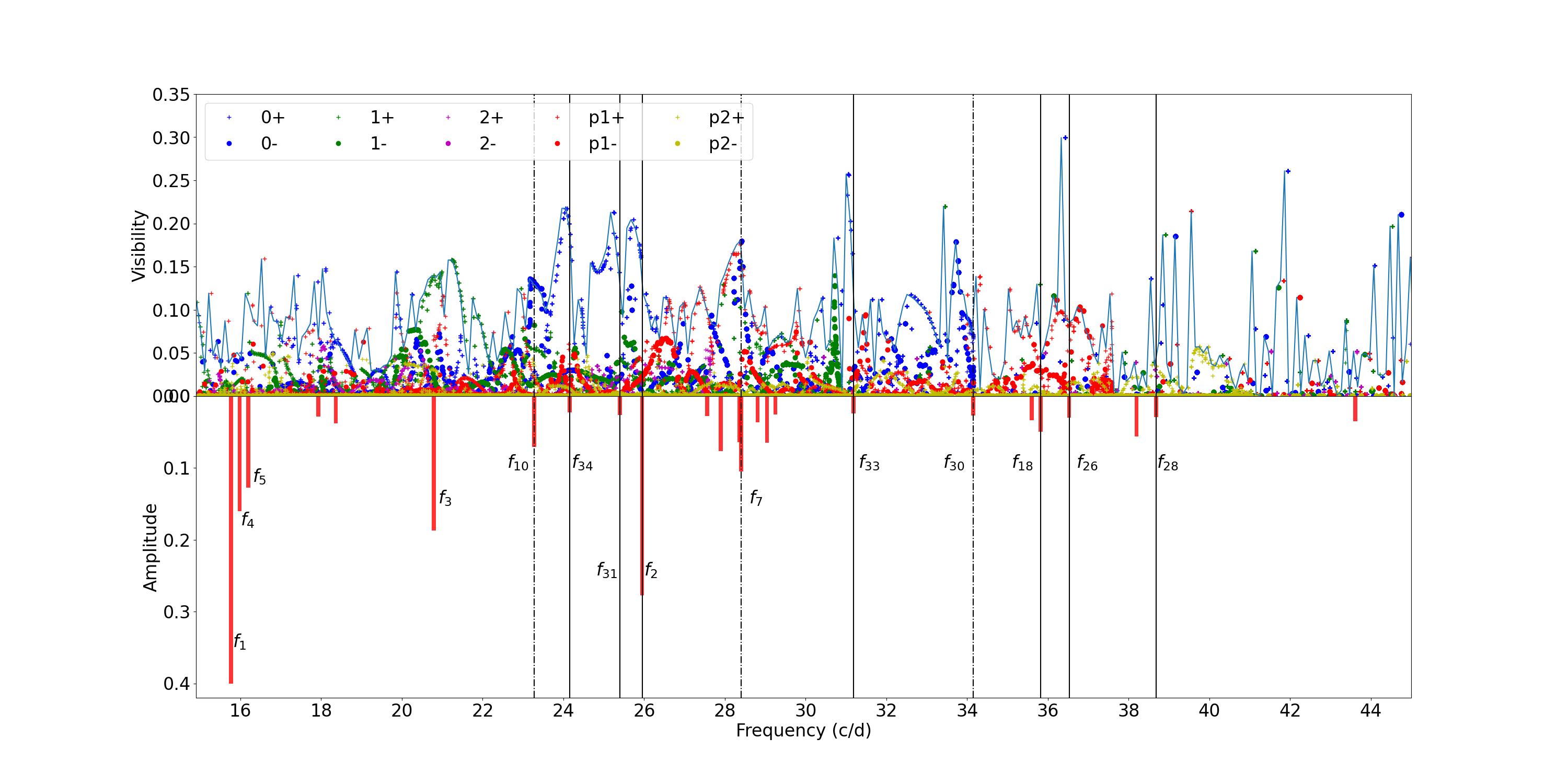}
\caption{Global view of eigenmodes visibility of Altair's concordance
model.{\it Upper part:} Proxy of visibility (see Eq.~\ref{formul_V}) as a function of frequency for all modes computed with the \topcode code using Altair's concordance model of \cite{bouchaud+20}. The symbols show the individual values for each mode. Pluses (dots) designate modes that are symmetric (anti-symmetric) with respect to the equator. Colours indicate the azimuthal symmetry (see the legend), and p1+, p1-, p2+, p2- indicate prograde modes (i.e. negative $m$'s). The solid blue line is the envelope of the symbols using bins of 0.1 c/d. It highlights modes with high visibilities. {\it Lower part:} Scaled amplitude (arbitrary unit) of observed frequencies (red bars). The thin vertical solid lines show possible identification of some $0^+$ modes. Thin vertical dot-dashed lines show the same for three $0^-$ modes. Part of the frequencies have been annotated to ease location.}
\label{visib}
\end{figure*}
\begin{figure}[t]
\centering
\includegraphics[width=1.0\linewidth]{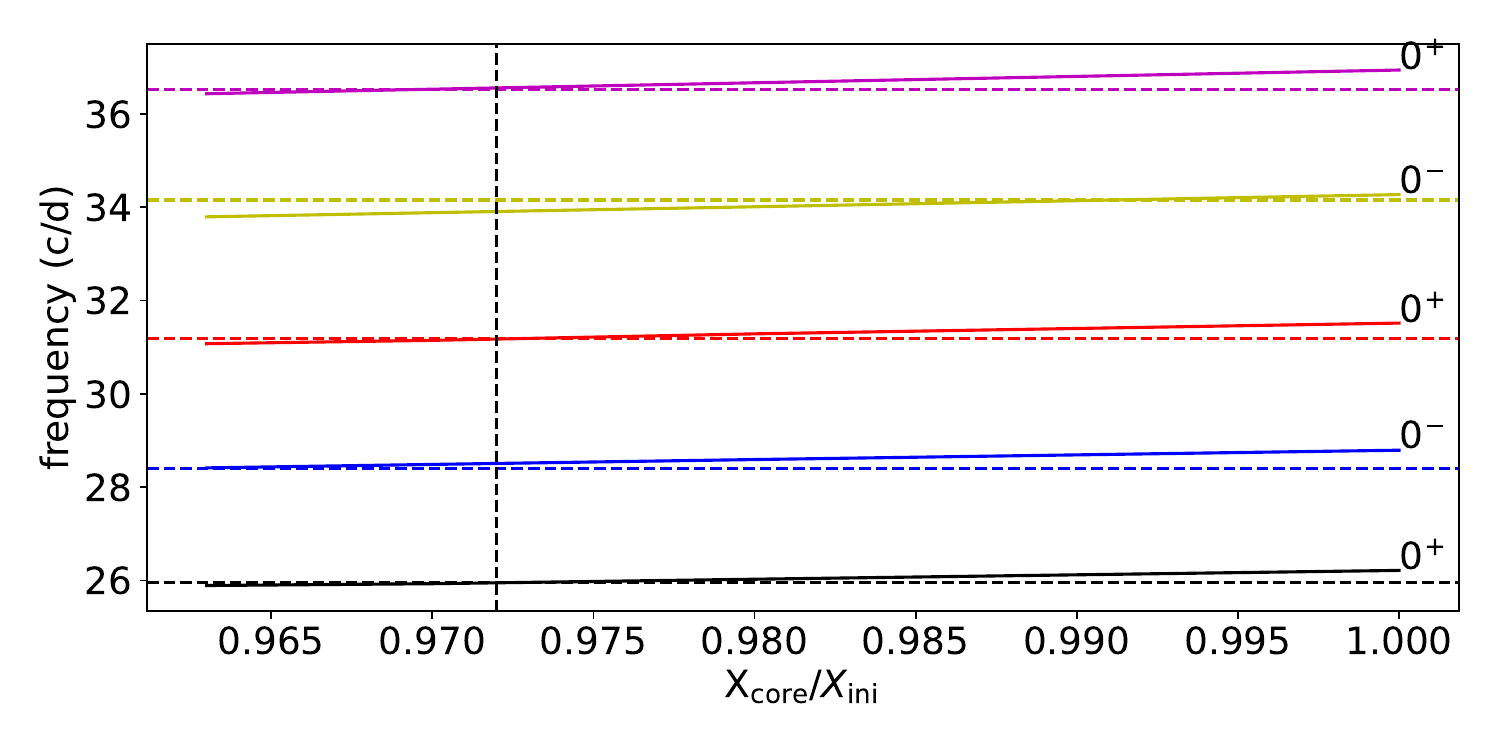}
\caption[]{Variation of the frequencies of five axisymmetric island
modes (solid lines) as a function of the hydrogen mass fraction in the
core with respect to that of the envelope. Here, $\rm X_{env}=X_{ini}=0.739$, as in \cite{bouchaud+20}. The horizontal dashed lines show the observed frequencies, while the vertical dashed line points out $X_c=0.972$, which is the value where the frequencies of three $0^+$ modes of the model match the observed ones.}
\label{fXc}
\end{figure}

We note that in the 2012 data set, \cite{ledizes+21} noticed a low-amplitude peak near 3.0 c/d, namely at the expected rotation frequency deduced from the concordance model of \cite{bouchaud+20}. No such peak appears in our data. If the 2012 detection is real, it may have been generated by a transient feature at the surface of Altair. A magnetic spot is then the most natural explanation since Altair is known to have some magnetic activity in X-rays \citep{robrade+09}.

Finally, we completed the monitoring of the amplitudes of the six most important modes by adding the measured amplitudes at the dates of observation, namely at 2022.555. The result is shown in Fig.~\ref{Ampl}, where one can see that the mode amplitudes of Altair are continuously varying over the years. {However, we emphasise that the foregoing variations may also be influenced by the difference in the bandpass of the instruments: WIRE observed in V+R \citep{buzasietal05}, MOST used the [375,675]~nm band \citep{walker+03}, and TESS uses the [600,1000]~nm band \citep{ricker+15}.

A tentative explanation for the amplitude variations given by \cite{ledizes+21}
is that modes are coupled to a thin subsurface convective layer where helium is twice ionised and where the kappa mechanism operates.
This coupling is a possible mechanism to induce time-varying amplitudes, but it needs numerical simulations to be confirmed and to show how it operates.

\begin{figure*}
    \centering
\includegraphics[width=1.\linewidth]{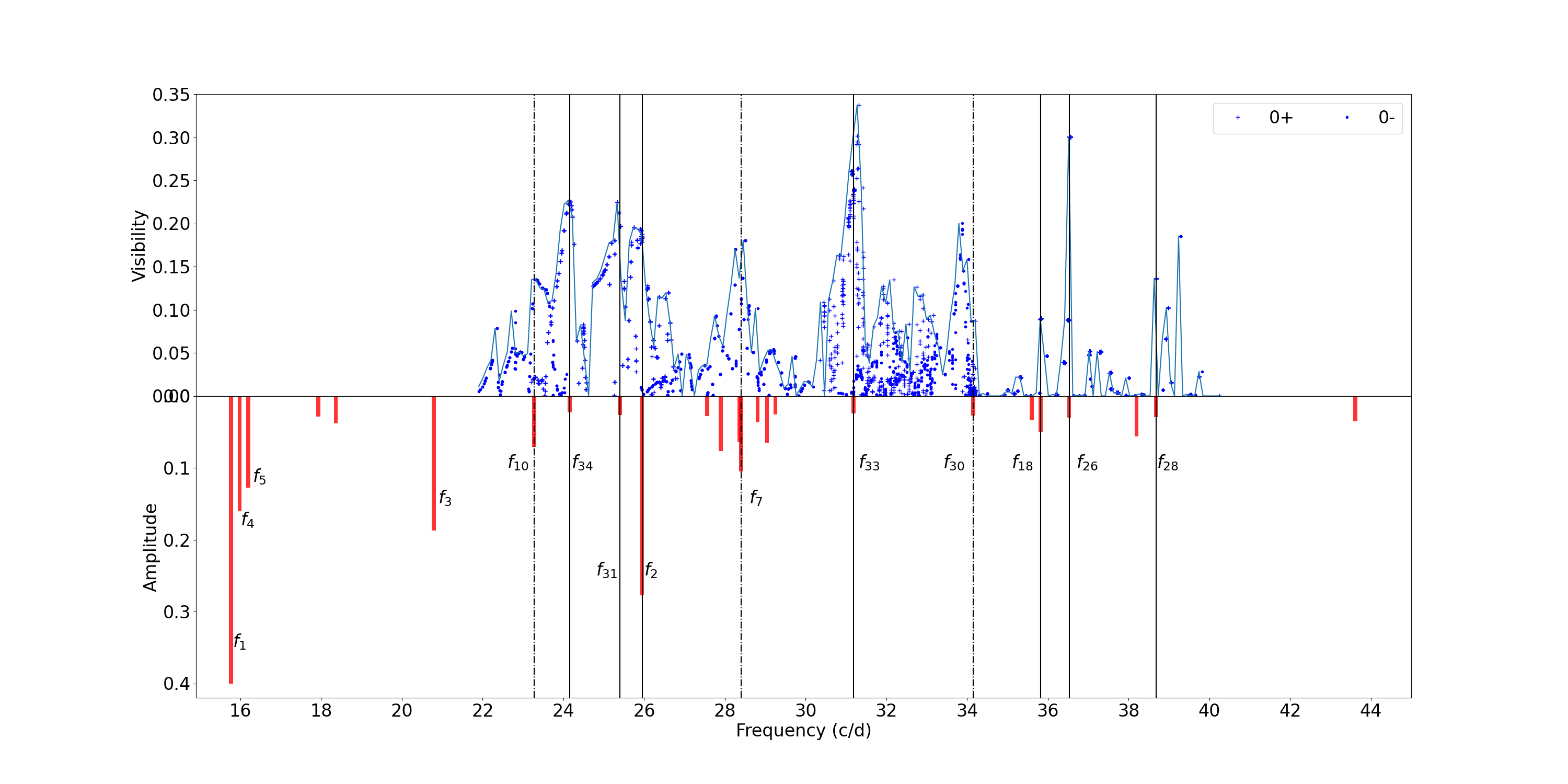}
\caption[]{Same as Fig.~\ref{visib} but only axisymmetric modes have been computed and plotted. Dashed-dotted vertical lines mark $0^-$ modes, while vertical solid lines highlight $0^+$ modes. The core hydrogen mass fraction has been slightly increased to $X_{\rm core}/X_{\rm env}=0.972$ to show the matching between modelled frequencies with high visibilities and observed frequencies.} 
\label{visinew}
\end{figure*}

\section{Forward modelling}

With this new set of frequencies we would like to know how models compare with these observations, especially the concordance model of \cite{bouchaud+20}, which was built to reproduce all data available in 2020. In particular, a few frequencies of the 1999 data set \citep{buzasietal05} could be retrieved. In this section, we make a larger and more complete investigation of the oscillation spectrum of Altair's concordance model and compare its predictions with our observations.

First, we recall the characteristics of the model of Bouchaud et al. It is a 2D
axisymmetric steady \ester model \citep{ELR13} with a mass of 1.863~\msun\ that rotates  with an equatorial velocity of V$_{\mathrm{eq}}$=313~km/s, thus at 74.4\% of the critical angular velocity. This high rotation rate induces a rotational flattening of 22\%. The metallicity of the model is Z=0.0192, and the hydrogen mass fraction is X$_{\rm ini}=0.739$ in the envelope and X$_{\rm core}/$X$_{\rm ini}=0.963$ in the core. This latter ratio was interpreted as corresponding to an age of $\sim100$~Myrs by \cite{bouchaud+20}.

As a first step in deciphering the eigenspectrum of Altair, we computed the adiabatic oscillations using the \topcode code, which can use a 2D-\ester model as an equilibrium model \citep{reese+21}.
Since the model is axisymmetric, eigenmodes are characterised by their azimuthal periodicity and equatorial symmetry, which we denote as $m^\pm$. Here, $m^+$ are modes that are symmetric with respect to the equator and whose azimuthal variation is $\propto\exp(im\varphi)$. Conversely, $m^-$ are equatorially anti-symmetric with the same $\exp(im\varphi)$ dependence.
Because TESS observations are photometric, we only expected $m=0,1,2$ to be visible. Indeed, in fast rotators such as Altair, eigenmodes are made of a long series of spherical harmonics strongly coupled in the $\ell$ index by the Coriolis acceleration, the centrifugal distortion of the background, and the differential rotation. However, since the model is axisymmetric, modes of different azimuthal wavenumber $m$ are uncoupled. Just as for the $\ell$ in a non-rotating star, only the low $m$ may provide modes with high visibility.
Hence, we focused on eigenmodes with the $m=0^\pm,1^\pm,2^\pm$ symmetries. For the non-axisymmetric ones, we included both retrograde and prograde modes, that is, $m$ can be positive or negative. 

The \topcode code determines eigenvalues using the Arnoldi-Chebyshev algorithm,
which is a Krylov-based method \citep{VRBF07} that allowed us to compute a small
number of eigenvalues around a given point of the complex plane. Since we
restricted ourselves to adiabatic oscillations, all eigenfrequencies are real.
We first computed eigenvalues around each observed frequency, and it revealed a
high density of frequencies, especially in the low-frequency
range.\footnote{This is not a surprise since gravito-inertial modes likely form a dense spectrum in the low-frequency range, as can be argued in some simplified set-ups \citep{DRV99}.} We therefore concentrated on frequencies larger than 15~c/d, which contain the oscillations with the highest amplitudes, and we scanned the frequency interval from 15 to 44~c/d.

\begin{figure*}[t]
\centering
\includegraphics[width=0.32\linewidth]{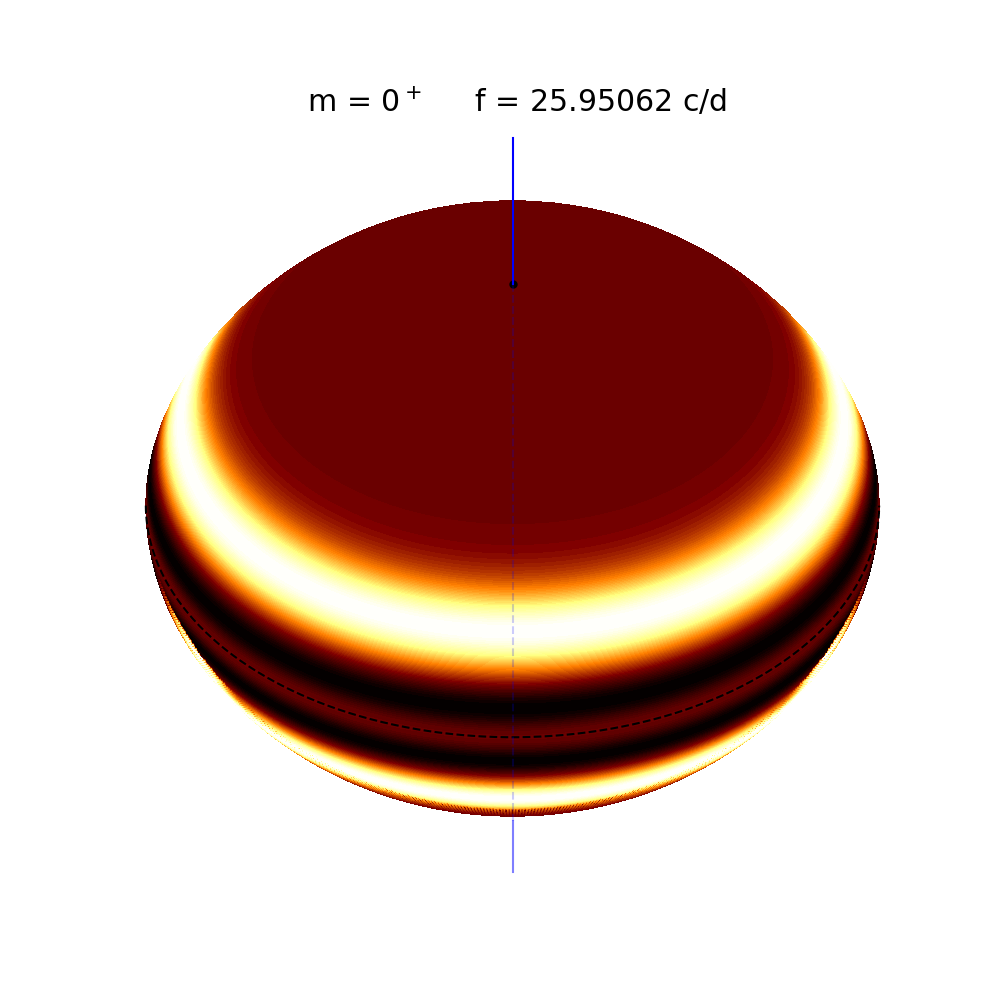}
\includegraphics[width=0.32\linewidth]{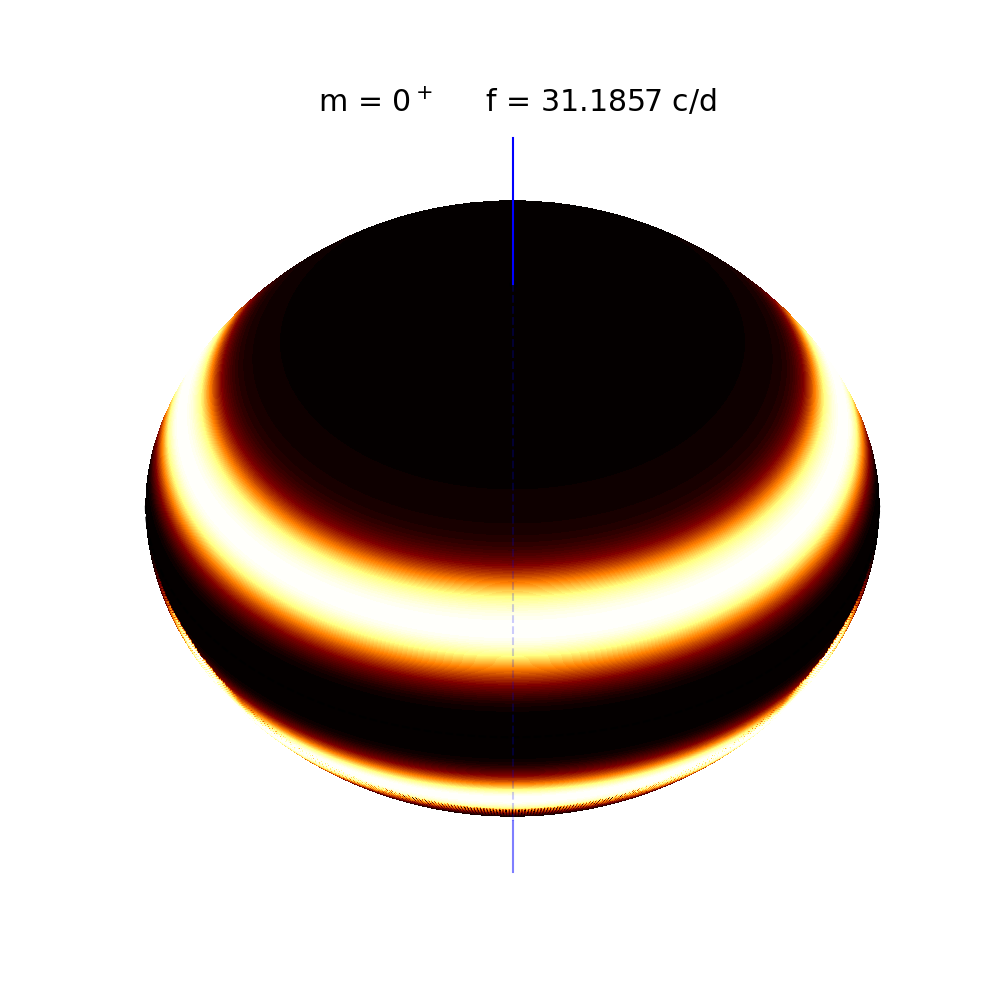}
\includegraphics[width=0.32\linewidth]{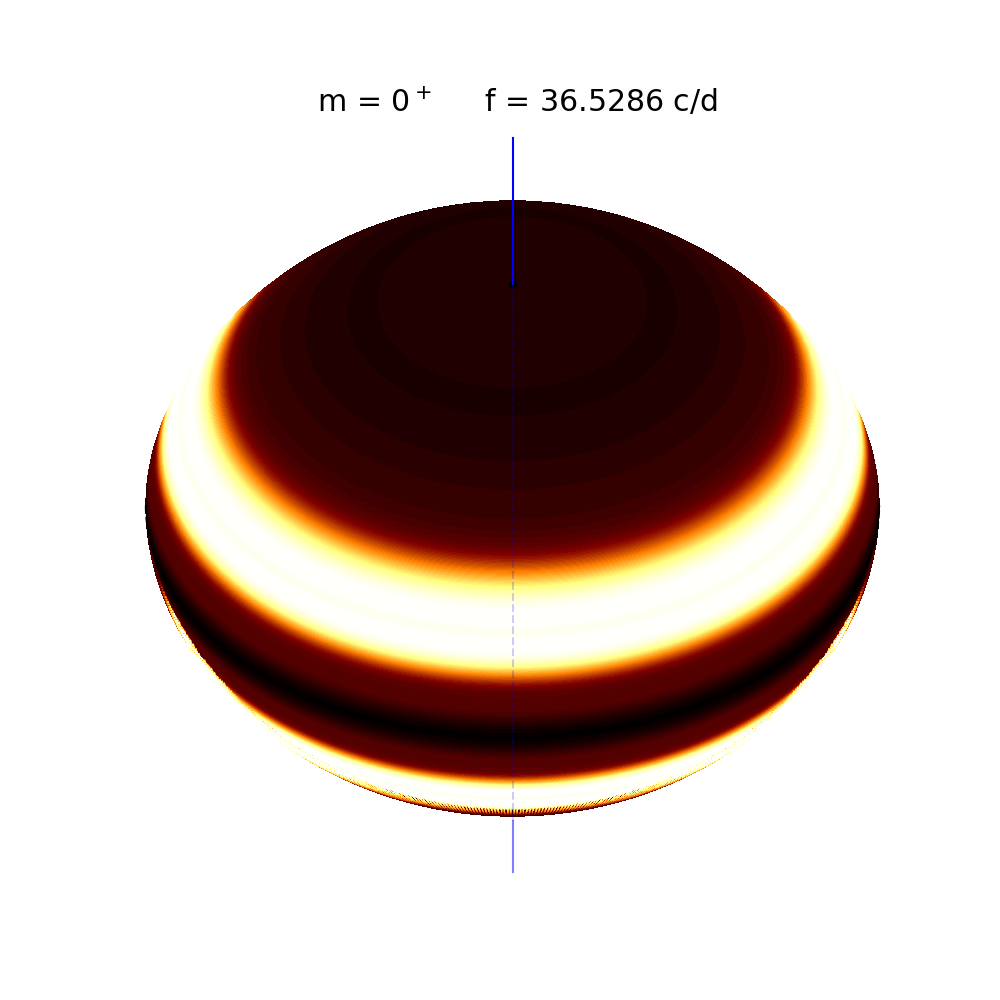}
\includegraphics[width=0.32\linewidth]{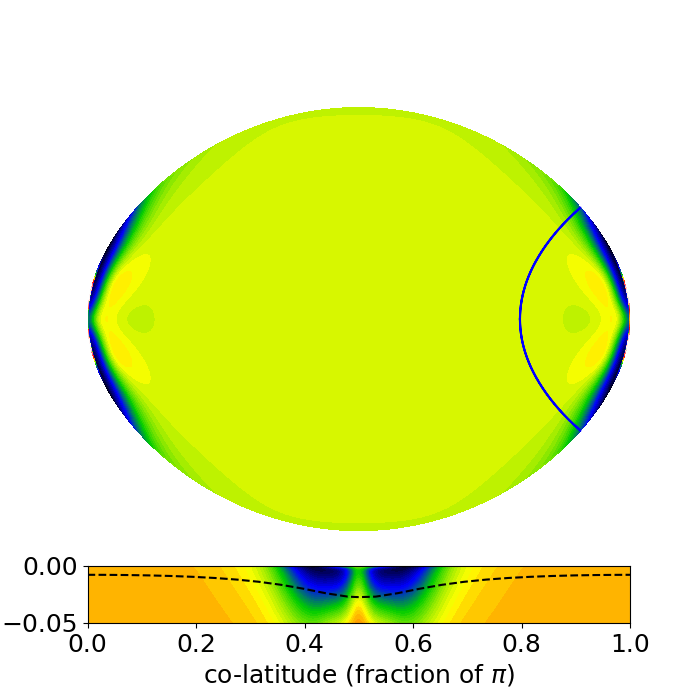}
\includegraphics[width=0.32\linewidth]{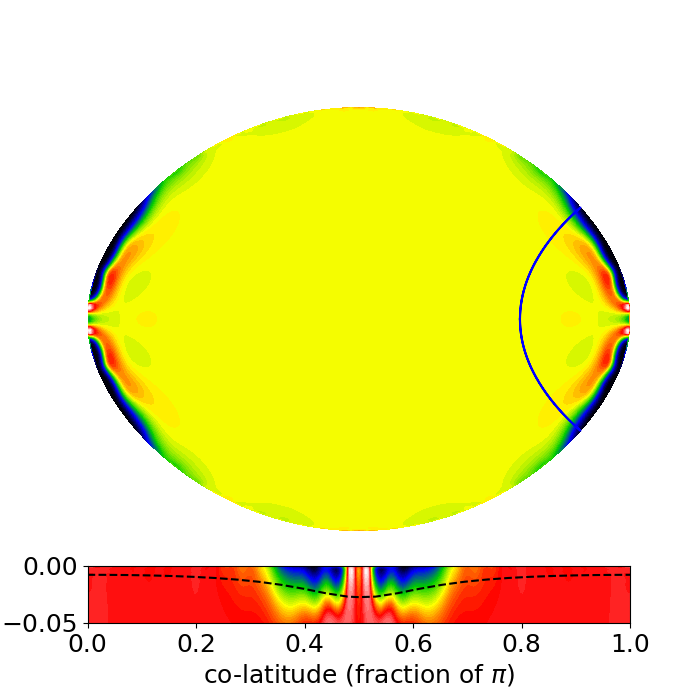}
\includegraphics[width=0.32\linewidth]{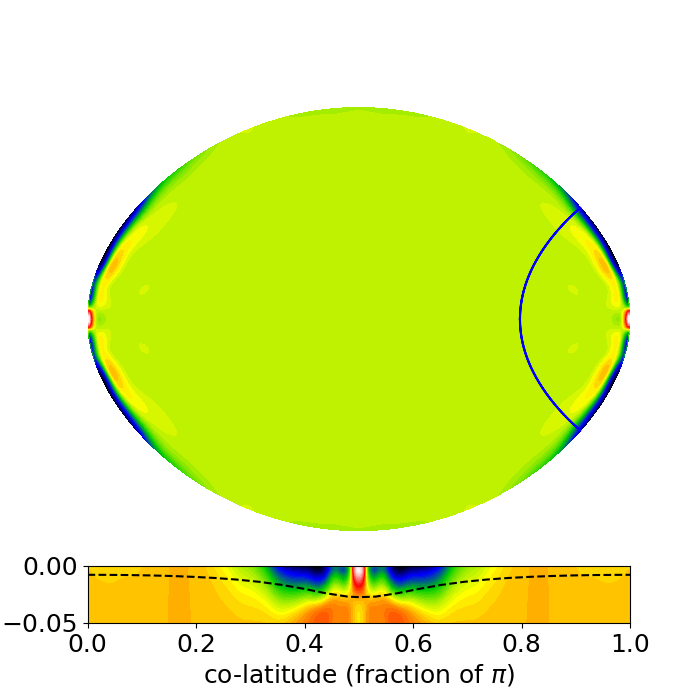}
\caption[]{Three $0^+$ modes identified in the spectrum of Altair. The top row shows the surface amplitude of the relative pressure fluctuation associated with the mode as seen from the Earth. The bottom row shows a meridional cut of the same quantity along with a zoom of the subsurface layers, with the depth being limited to 5\% of the equatorial radius. The solid line in the meridional cut of the star shows the trajectory of an acoustic ray associated with the mode. The dashed black line in the zoomed layers highlights the 50,000~K isotherm around which the second ionisation of helium takes place.}
\label{views}
\end{figure*}

Interferometric observations have determined quite precisely the
inclination of the rotation axis of Altair with respect to the line
of sight (see introduction). Hence, we could compute the visibility
of each mode and search for the most visible modes.
As we had restricted ourselves to adiabatic oscillations, we had no
predictive power over the mode amplitudes. Yet, we assumed that those with
the largest scales are the least damped or the most unstable and thus
the most visible as well. {Noting that the true visibility is tightly
related to the temperature fluctuation, which in an adiabatic model is
itself strongly related to the pressure fluctuation,} we defined a proxy
of the visibility as

\begin{equation}
    \rm V_{\rm proxy} = \frac{\int_{S_v} (\delta p /P) \vu\cdot\dS}{\max_S(|\delta p /P|)}
    \label{formul_V},
\end{equation}
where $S_v$ is the visible part of the stellar surface, $\vu$ is a unit vector pointing towards the observer, and $\dS=\vn dS$ is the surface element directed along the normal of the surface. 
This proxy of the visibility is faster to compute than the true visibility of a mode as described in \cite{reese+13}, and it still allowed us to qualitatively compare the visibility of different modes and to display the most visible ones. 

The resulting plot of the foregoing proxy is shown in Fig.~\ref{visib}. In this figure (upper part), each symbol refers to an eigenfrequency of \cite{bouchaud+20} concordance model. One may note the large density of eigenvalues in any interval of frequencies. As the $y$-axis shows the visibility of the modes, the peaks of the upper envelope of the symbols (the blue line) point out frequencies associated with eigenmodes with only large-scale variations on the surface of the star. In the lower part of  Fig.~\ref{visib}, we have drawn red bars representing the observed oscillation amplitudes (arbitrary unit) at the observed frequencies. Thin vertical lines show possible identifications between visibility peaks and observed frequencies. We note that all of these possible identifications are associated with axisymmetric modes (blue symbols). Only $f_3$ (at 20.8~c/d) seems to be associated with a non-axisymmetric $1^+$ retrograde mode. Surprisingly, the most prominent observed mode, at $f_1=15.77$~c/d, is not associated with any highly visible mode of the model. We discuss this point below.

Regarding Fig.~\ref{visib}, we also note that the model frequencies
(peaks of visibilities) are systematically lower than the observed
frequencies. If these frequencies are associated with acoustic modes,
as suggested by \cite{bouchaud+20}, this shift means that the model
is not dense enough since acoustic frequencies scale with the average
density of the star \citep{reese+08,garciahernandez+15}. In other words,
the concordance model of \cite{bouchaud+20} is slightly too old. For five
axisymmetric modes, which can be associated with observed frequencies,
we examined the frequency dependence with $X_{\rm core}/X_{\rm ini}$,
namely with the hydrogen mass fraction in the core scaled by its initial
value. This dependence is illustrated in Fig.~\ref{fXc}. We note that
the $0^+$ modes, which are axisymmetric and symmetric with respect to
equator, all point to a ratio of $X_{\rm core}/X_{\rm ini}\simeq0.972$,
while the concordance model used $X_{\rm core}/X_{\rm ini}=0.963$. We
also note that $0^-$ modes (antisymmetric with respect to equator)
do not agree with the $X_{\rm core}/X_{\rm ini}$ value pointed out by
the $0^+$ modes. The $0^-$ modes even disagree among
themselves: $f_7$ suggests a ratio of $\sim0.963$, while $f_{30}$ points
to $\sim0.99$. A misidentification of the associated modes could explain
the different $X_{\rm core}$ inferred from the $0^-$ modes,
but this may also reveal an unrealistic profile of the
sound speed along the acoustic rays. The error in the sound speed may
be compensated in symmetric modes but emphasised in anti-symmetric
ones. Further investigation is needed.

The previous remarks prompted us to recompute eigenfrequencies of a model with $X_{\rm core}/X_{\rm ini}=0.972$. However, in this case we restricted our calculations to axisymmetric modes and frequencies between 22 and 39 c/d so as to gain clarity and save computation time. The results are shown in Fig.~\ref{visinew}. As expected, the matching of $0^+$ frequencies is quite good, and that of $0^-$ frequencies is fair enough.

The foregoing results extend the previous possible identifications of frequencies of Altair's spectrum. Indeed, \cite{bouchaud+20} identified the frequencies [23.2791, 25.95061, 28.4064] c/d with three axisymmetric modes of symmetry $[0^-, 0^+, 0^-]$, respectively. We extend this series with six additional modes at frequencies [24.1551, 25.4012, 31.1858, 34.1521, 35.8214, 36.5285] c/d of symmetries $[0^+,0^+,0^+,0^-,0^+,0^+]$, respectively. We acknowledge that the identification of frequency $f_{30}$ is only tentative.

All foregoing modes are acoustic modes, so-called island modes in the terminology of \cite{LG09}. They are associated with periodic orbits of acoustic rays bouncing between low north and south latitudes, as illustrated in Fig.~\ref{views} for three symmetric $0^+$ modes at $f_2$, $f_{26}$, and $f_{33}$.

As noted previously, the most prominent observed frequency at
15.76789~c/d does not seem to be associated with any highly visible mode
of the model. However, this value is reminiscent of the frequency of the
fundamental mode of $\delta$-Scuti stars, as can be derived from the
period-luminosity relation given by \cite{barac+22}. This relation
points to a frequency of $\sim$16~c/d for Altair. However, Altair is
young (see below) and rotating extremely rapidly ($\nu_{\rm
rot}\simeq3$~c/d). To find the fundamental mode of Altair, we therefore
reduced the angular velocity of the model to 15\% of the critical
angular velocity (instead of its actual 74.4\%) while keeping all other
parameters constant. As a result, we found the fundamental mode at a
frequency of 20.88~c/d. This value, which is higher than the
period-luminosity relation prediction, is due to the quasi-ZAMS nature
of our model. We then increased the rotation rate of the model,
following the evolution of the fundamental mode. As expected, its
frequency decreased due to the volume increase of the model. However,
when the rotation rate reached 50\% of the critical angular frequency,
we could no longer follow the mode, as it mixes with gravito-inertial
modes at this point. Computations of eigenvalues around the expected
value of the fundamental frequency showed many avoided crossings,
making identification of the fundamental mode extremely difficult. An
impression of the complexity of the spectrum in this frequency range is
given by Fig.~\ref{visib}, but it can also be appreciated with Fig.~4
of \cite{RLR06}. To decipher the oscillation properties of Altair around
its dominant oscillation frequency at 15.77~c/d, a dedicated study will
be needed that likely includes non-adiabatic and non-linear effects.

\section{The age of Altair} 

The foregoing possible identification of eight eigenfrequencies of Altair,
implying $X_{\rm core}/X_{\rm ini}\simeq0.972$, prompted us to improve the
estimated age of the star, which was calculated to be $100$~Myrs by
\cite{bouchaud+20}. To this end, we used a new version of the \ester code that
can compute time evolution of a 2D model \citep{mombarg+23,mombarg+24a}. This
new version of the code can tell us how much time is needed to reach the
inferred hydrogen mass fraction in the core, but presently it cannot compute the
pre-main sequence phase of the model. To estimate this latter phase, we reverted
to MESA models
\citep{paxton+11,paxton+13,paxton+15,paxton+18,paxton+19,jermyn+23}. With these
1D models, we evaluated at 26~Myrs, the duration of the pre-main sequence for a
model of 1.863~\msun\ and a similar chemical composition as the \ester model. The appropriate evolutionary 2D-\ester model says that reducing $X_{\rm core}/X_{\rm ini}$ from unity to 0.972 takes 62~Myrs. Hence, we estimate the age of Altair to 88~Myrs. For further illustration, we give in Table~\ref{ages}, the ages given by the MESA code for a non-rotating model and a rotating model. For this latter model, we set the rotation rate to about 65\% of the critical one. This rotation is the highest acceptable by the code. Although this value is clearly out of the acceptable ones for a 1D model, it gives an idea of the spherically averaged effects of rotation, which turn out to be quite modest. From the numbers shown in Table~\ref{ages}, we observed that MESA models evolve more rapidly than the 2D-\ester model, essentially because of a slightly higher central temperature. The origin of the age differences between the models is likely to be found in the microphysics (e.g. opacities, equation of state, nuclear reaction rates). The dimensionality of the model, 1D or 2D, is crucial to estimating the fundamental parameters of the star (mass, radii, rotation rate, for instance) but is no longer crucial for time evolution in the case of Altair since this star is very young and secular effects of rotation have not had time to accumulate. Our conclusion is that in the present state of stellar models, the age of Altair is close to 90~Myrs with an uncertainty we estimate to 10~Myrs due to the differences in microphysics between the models. Finally, we note that the slight increase of $X_{\rm core}/X_{\rm ini}$ changes the size of the star by a negligible amount compared to the uncertainties of interferometric radii measurements.

\begin{table}
\caption{Altair's age given by the models.}
\centering
\begin{tabular}{|c|c|c|c|}
 $X_{\rm core}/X_{\rm ini}$ & \MESA & \MESA-rot &  \ester-2D \\
   0.972  & 77 Myrs & 79 Myrs &  88 Myrs \\
   0.963 & 99 Myrs & 101 Myrs &  108 Myr \\
\end{tabular}
\label{ages}
\end{table}

\section{Summary and conclusions}

Using the halo photometry technique \citep{white+17}, we processed the TESS images of Altair and derived its light curve during the observation period from 10 July to 5 August 2022. Using the \FELIX code \citep{charpinet2010,zong2016}, we derived the power spectrum of the light curve and exhibited 34 frequencies. Compared to previous works \citep{buzasietal05,ledizes+21} that revealed 15 frequencies, we confirmed 12 of them, and we add 22 new frequencies thanks to the high quality of the data. Previous data came from the MOST and WIRE satellites but were plagued by scattered light of the Earth due to the low-altitude orbits of their missions. The amplitudes of the repeatedly observed modes still show variations, which we tentatively attribute to the probable coupling of the modes with a thin convective layer near the surface of Altair \citep{ledizes+21}.

We then focused on a subset of frequencies that were previously identified with those of axisymmetric modes during the making of a concordance model of Altair by \cite{bouchaud+20}. The identification of modes at the frequencies $[23.2791,25.95061,28.4064]$ c/d by this previous work appears reliable. By knowing the inclination of the rotation axis with respect to the line of sight, we could compute the visibilities (in fact a proxy of it) of all modes, which we determined using the concordance 2D model of \cite{bouchaud+20} and the \topcode code for the oscillations \citep{reese+21}. We thus identified seven new frequencies still associated with axisymmetric modes.

Among these ten candidates, we selected three modes that are symmetric with respect to the equator and used their frequencies to better determine the size of the star, which meant determining its age, as we assumed the mass given by \cite{bouchaud+20}. The three frequencies indeed point to the same hydrogen mass fraction in the core, namely $X_{\rm core}/X_{\rm ini}\simeq0.972$, with $X_{\rm ini}=0.739$ \citep{bouchaud+20}. Re-computing the concordance model with this slightly higher hydrogen mass fraction in the core led to a nice matching between nine observed frequencies and nine modelled frequencies. With time-evolving 2D-\ester models, we could then estimate our best model age, namely 88~Myrs, which is slightly younger than estimated by the previous concordance model of \cite{bouchaud+20}, who found $X_{\rm core}/X_{\rm ini}=0.963$  and which corresponds to an age of 108~Myrs. Nevertheless, we still estimate the uncertainty of the age to be about 10~Myrs based on comparing the \ester and \MESA predictions. We argue that the microphysics are the main source of uncertainty in the evolution of the hydrogen mass fraction of the core, rather than the dimensionality of the model. Improvements in this direction may help us reproduce more observed frequencies with the models.

The foregoing results show that the modelling of Altair is
progressing. However, many frequencies still have to be identified,
and it remains a challenge to determine the origin of the oscillation
at 15.76789 c/d, which is the most prominent and has the highest
amplitude. Around this frequency, no highly visible mode pops out of
the model. We investigated the change of frequency of the fundamental
mode of a model of similar mass but of lower rotation and we found that
such a fundamental mode disappears when rotation is increased beyond 50\%
of the critical one. A series of avoided crossings with (presumably)
gravito-inertial modes causes the fundamental acoustic mode to mix
with these modes, whose spectral density is high.  It is very likely
that this prominent oscillation results from a combination of several
eigenmodes and requires a non-adiabatic and non-linear approach to be
understood. Investigation of the non-linear dynamics of all the detected
modes remains a highly desired future step.

\begin{acknowledgements}
We are very grateful to the referee  for many valuable comments, which helped us improve the original manuscript. MR, DRR, and JSGM gratefully acknowledge the support of the French Agence Nationale de la Recherche (ANR), under grant MASSIF (ANR-21-CE31-0018). MR and SC acknowledge support from the Centre National d'Etudes Spatiales (CNES). MR also acknowledges support from the European Research Council (ERC) under the Horizon Europe programme (Synergy Grant agreement N$^\circ$101071505: 4D-STAR).  
Computations of \ester 2D models and the oscillation spectra have been possible thanks to HPC resources from CALMIP
supercomputing center (Grant 2023-P0107). While partially funded by the European Union, views and opinions expressed are however those of the authors only and do not necessarily reflect those of the European Union or the European Research Council. Neither the European Union nor the granting authority can be held responsible for them.
\end{acknowledgements}

\bibliographystyle{aa}
\bibliography{bibnew}

\end{document}